# Comment on "How many principles does it take to change a light bulb…into a laser?"

**Wolfgang Elsäßer**

Institute of Applied Physics, Technische Universität Darmstadt, Schlossgartenstrasse 7, 64289 Darmstadt, Germany

E-mail: Elsaesser@physik.tu-darmstadt.de

**Abstract**

In a recent article entitled "How many principles does it take to change a light bulb…into a laser?" by Howard M. Wiseman, Physica Scripta 91 (3), 033001 (2016) the author addresses the question as expressed by the title in summarizing what he considers as the fundamental features which distinguish laser light from thermal light. He draws the conclusion that there are in fact 4 principles making this difference which he outlines in detail. Here, in this comment I am analyzing this extremely well formulated question in the framework of the Hanbury-Brown & Twiss experiment, thus also celebrating its 60$^{th}$ anniversary. I am giving a straightforward, concise and unique answer to this question, namely emphasizing that the intensity correlations or the photon statistics make the difference between light emitted by a light bulb and that of a laser.

___

I am indebted to the author for this very nicely chosen title in formulating very elegantly the question "How many principles does it take to change a light bulb…into a laser?" [1] and I am very grateful for his efforts to illuminate this issue. At a first glance it appears to be a simple question, which in fact is very often wrongly answered even by final year Bachelor students. Indeed, we have to work hard to finally set them on the right track when it comes to a correct understanding of the differences between the properties of a laser and those of a light bulb [2]. However, while appreciating the author's credits, I am equally disappointed by his final answer. In fact, 2015 has been the Unesco International Year of Light (IYL) and in 2016, we are celebrating a milestone in the field of quantum optics, the 60[th] anniversary of the Hanbury-Brown & Twiss (HBT) experiment, at least in its terrestrial optical version in 1956 [3]. This experiment has paved the fundaments of quantum optics even before the lasers existed [4]. I am wondering why the outcome of this fundamental experiment in the spirit of quantum optics even today at the occasion of the 60[th] anniversary with a tremendous outreach and impact far beyond optics, e.g. into nuclear physics, solid state physics and quantum gases is not exploited by the author in order to elaborate and analyze the difference between these two antipodes of a light source.

This HBT experiment gives a simple, elegant, clear and unique answer on the title question, namely the intensity correlations or the photon distribution function P(n) make the difference between light emitted by a light bulb or by a laser. Photon bunching in the picture of enhanced intensity correlations for thermal light showing a central second order correlation value $g^{(2)}(\tau=0)$ value of two is confronted with that of a value of unity for laser light. This result is involved with the observation of Bose-Einstein statistics or Poissonian statistics for the thermal light bulb and the laser, respectively [5, 6, 7]. Furthermore, non-classical light, either in the regime below the $g^{(2)}(\tau=0)$ border line with sub-Poissonian statistics and thus squeezing [8] or even above the $g^{(2)}(\tau=0) =2$ value equivalent to super-bunching [9] is fully captured in this picture of intensity correlations. Part of this experimental progress alongside with the parallel developed understanding was recently only possible by exceeding the measurement time resolution limit of determining photon correlations, e.g. by exploiting ultra-fast two-photon absorption (TPA) multiplier techniques in this field [10]. All these results of intensity correlations make it obvious that only considering the spectral properties in terms of the first order (field) correlation $g^{(1)}(\tau=0)$ obtained via the Wiener-Khintchine theorem is no longer a criterion for differentiating between thermal and laser light [11 ,12,13].

Also directionality is no longer a unique criterion for laser light because amplified spontaneous emission originating from semiconductor-based optoelectronic light emitters with waveguides unifies broad-band and directionality and does exhibit photon bunching, i.e. thermal light second order coherence characteristics [14]. These ultra-compact well-controlled sources even allow for coherence tailoring when applying well-controlled optical feedback such that a continuous coherence transition is achieved [15] consistent with a multimode, phase-randomized Gaussian (PRAG) quantum state model [16]. These experiments demonstrate and visualize that optical feedback can be exploited to achieve a continuous transition between the 1[st] and 2[nd] order coherence properties of a thermal emitter and a laser [15], thus being fully in-line with the elegantly formulated title question. Recently, it has been demonstrated experimentally that with broad-area superluminescent diodes (BA-SLDs) fully incoherent light can be generated, light being firstly incoherent in first order coherence, i.e. broad-band in the spectrum, secondly spatially incoherent due to the multi-filament multi-mode output at the 200µm wide facet, and finally and thirdly incoherent in $2^{nd}$ order, i.e. being a thermal light source with photon bunching and Bose-Einstein statistics [17]. This has stimulated excitement in the scientific community, in particular when applying these newly-considered light sources as an ultra-compact classical light source [17, 18] for classical ghost imaging [19, 20].

Therefore, based on the reasons outlined above, I am convinced that when considering the Hanbury-Brown & Twiss intensity autocorrelation facts, higher order correlations or photon statistics criteria, the highly interesting and challenging question of the title can be answered quite straightforwardly, uniquely and convincing.

In conclusion, all this together shows that in fact it needs only one more principle, namely considering intensity correlation function and/or photon statistics features, not to change a light bulb into a laser but to understand the difference between them.

**Acknowledgements**

I would like to thank Dr. Martin Blazek and Sébastien Hartmann for numerous stimulating discussions in the framework of this topic, thus having perpetually developed my point of view of coherence and particularly of the HBT experiment. Part of this work has been supported by the Deutsche Forschungsgemeinschaft (DFG: project EL 105/21).